# Commissioning of CMS zero degree calorimeter using LHC beam

**O Grachov[1], M Murray, J Wood**
Department of Physics and Astronomy, University of Kansas
Lawrence, KS USA
**Y Onel, S Sen, T Yetkin**
Department of Physics and Astronomy, University of Iowa
Iowa City, IA USA
On behalf of HCAL-CMS collaboration

E-mail: grachov@ku.edu

**Abstract.** This paper reports on the commissioning and first running experience of the CMS Zero Degree Calorimeters during December 2009. All channels worked correctly. The ZDCs were timed into the data acquisition system using beam splash events. These data also allowed us to make a first estimate of channel-by-channel variations in gain.

## 1. Introduction
The CMS Zero Degree Calorimeters (ZDCs) [1] are located at the straight section of the LHC, between the two beam pipes at 140 m on each side of the interaction region. They are sampling calorimeters using tungsten (W), as radiator and quartz fibers (QF), as the active media. A significant advantage of this technology is that the calorimeter can be very compact, extremely fast and radiation hard. The Cherenkov light generated by charge particles passing through the fibers is brought to photomultiplier tubes (PMT). Schematic illustration of the ZDC is presented in Figure 1. Each ZDC includes electromagnetic (EM) and hadronic (HAD) section. Each section consists of a tungsten plate/quartz fiber ribbon stack. The hadronic section is longitudinally segmented into 4 readout towers. The electromagnetic section segmented into 5 transverse towers with width of 15.86 mm. Both sections are 79.3 mm wide and 100 mm high. The main parameters of ZDC are presented in Table 1. The ZDC is fully functional and fully integrated into CMS [2] operations. Following very extensive local and global CMS running, commissioning was carried out first with splash events created by the circulating proton beams at 450 GeV impinging on a target. These splash events were useful for adjustment of read-out timing and to determine the relative timing of ZDC channels, as well as to validate relative gain calibration of the detector's towers.

---

[1] To whom any correspondence should be addressed.

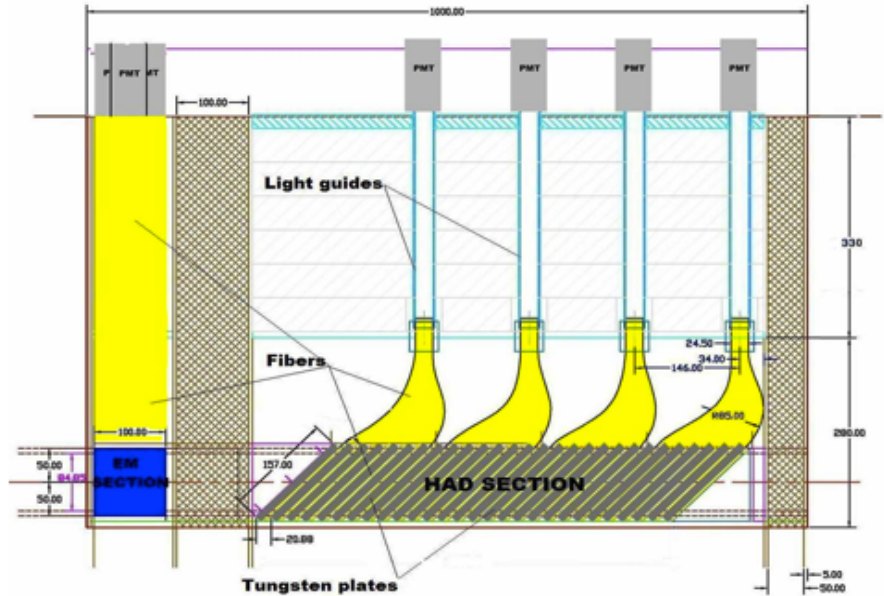

**Figure 1.** Schematic illustration of the ZDC (EM section and HAD section).

Table 1. Parameters of the ZDC

|                          | Hadronic Section        | Electromagnetic Section |
|--------------------------|-------------------------|-------------------------|
| Sampling Ratio           | 15.5 mm W/0.7 mm QF     | 2 mm W/0.7 mm QF        |
| Number of Cells          | 24                      | 33                      |
| Interaction (Rad.) Length| 5.6$\lambda$o           | ~1$\lambda$o(19Xo)      |
| Number of Channels       | 4 longitudinal segments | 5 transverse towers     |

## 2. Detector's commissioning

### 2.1. Calibration with test beams

ZDC detectors were constructed at University of Kansas and assembled at CERN, and prior to installation were tested at CERN SPS test beam in 2006 and 2007 [3, 4]. The test beam line delivered positrons with energy between 10 GeV and 150 GeV and positive pions with energy 150 GeV, 300 GeV and 350 GeV. The energy resolution for positrons is $(\sigma/E)^2$ = 0.49/E+ 0.0064. The linearity of the response as a function of positron beam momentum has been measured and the calorimeter is found to be linear over a range of from 10 GeV to 150 GeV to within 2%-3%. The energy resolution of the combined calorimeter (EM+HAD) for hadrons was obtained by a Landau fit and it is 21.5 % for 300 GeV positive pions. An extrapolation to energy 2.75 TeV will give the resolution of about 15%. Measurements show good linearity of detector in the positive pion energy range of 150GeV to 350 GeV.

### 2.2. Beam splash measurements

A total of 570 beam splash events for ZDC plus and 150 events for ZDC minus were recorded during the LHC 2009 commissioning exercises. The LHC beam with energy 450 GeV was steered into collimators located 150 m upstream in either direction of the CMS detector interaction point. Because of the large amount of material in the collimator only muons were able to penetrate the entire ZDC detector and form the signal in the calorimeter towers. A schematic representation of the geometry of the beam splash setup is shown in Figure 2.

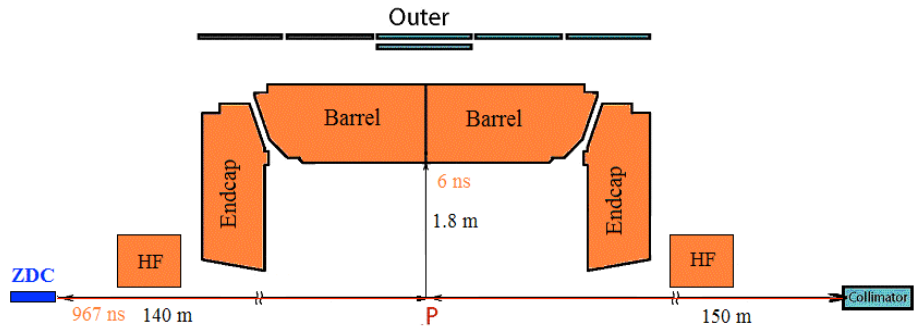

**Figure 2.** Beam splash measurement setup for one ZDC, showing the location of outer, barrel, two end caps and two HF hadron calorimeters, and ZDC calorimeter as well. ZDC located 140 m from CMS interaction point and 290 m from upstream collimator.

The splash events could not be used to study the absolute energy scale calibration because muons are minimum ionizing particles. Splash events were helpful for measuring the relative gain and delay between different channels of each ZDC and for overall adjustment of the timing of the calorimeter read-out.

*2.3. Read-out*
The read-out for one ZDC channel is shown in Figure 3. This scheme is the same as for read-out of HF (forward hadron) calorimeter [5].

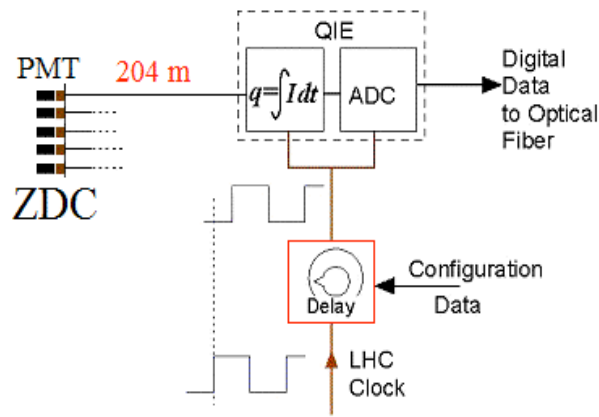

**Figure 3.** The read-out for one ZDC channel.

We are using 204 m length signal cable between phototubes and QIE (Q-charge, I-integrated, E-electronics). The QIE is a custom chip that contains the charge integrating electronics with an analog - to digital converter (ADC). The electric current collected from the PMT is integrated over each LHC clock period and then sampled. The integration clock can be delayed with respect to the LHC clock by programmable settings, which have a resolution of 1 ns.

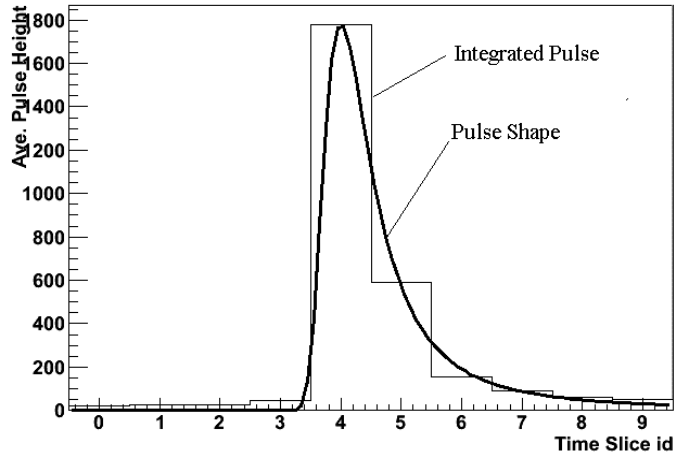

**Figure 4.** Integrated pulse and pulse shape. One time slice is 25 ns. The signal pulse is expected to start from the fourth sample and the baseline value can be estimated from the first three digitized samples (time slices 4,5 and 6).

Before the first beam, splash event's sampling delays were estimated from dead reckoning (time of flight, length of signal cable, electronic delay) and loaded into the ZDC front end electronics. These delay values were calculated to synchronize the ZDC with the rest of CMS. When beams arrived, events were selected by the CMS triggers system. For each event selected by the trigger the ten samples corresponding to the event were transmitted to the electronics for insertion into the CMS data stream. In order to estimate the amplitude of the signal and the timing for each channel, a fit was performed to the 10 digitized 25 ns time samples. The timing of each channel was calculated using charge-weighted algorithm based on two time slices around time slice with max energy deposition. The delay of the read-out pipeline is such that the signal pulse is expected to start from the fourth sample and the baseline value can be estimated from the first three digitized samples (time slices 4,5 and 6) (see Figure 4).

## 3. Results of measurement

*3.1. Timing*
The resulting distribution of the difference between the corrected time (after first splash beam) and predicted dead reckoning time is shown in Figure 5. This difference was corrected. Additional possible systematic effects, such as an offset in timing between the positive and negative ZDC, were studied with beam splash data from both directions. The timing for the ZDC plus before any corrections shows an average time differences that is 16 ns later than ZDC minus (see Figure 6). We corrected this difference and checked it using additional beam splash measurements.
The ZDC channel-by-channel synchronization was verified. Time variation is in the region of ±6 ns. For each individual EM and HAD channel, the signals generated by particles originating from the interaction point (IP) are registered with the different time, because of their different flight times and different intrinsic delays. The final off line time corrections are determined with collision data and expected to provide synchronization with a spread in the per channel mean times of 1 ns.

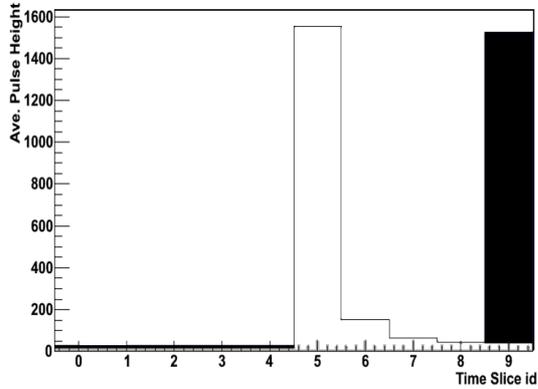
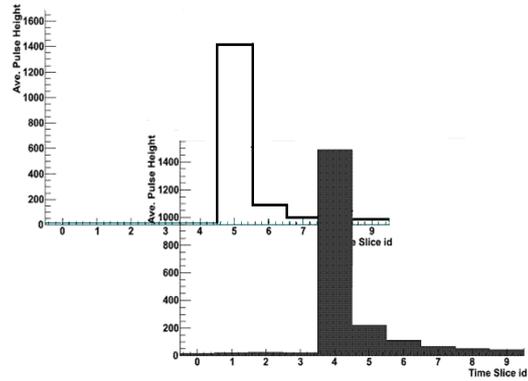

**Figure 5.** The resulting distribution of the difference between the corrected time (white) and predicted dead reckoning time (black) for positive ZDC. Data from LHC beam commissioning were used to validate the ZDC sampling delay settings derived from dead reckoning estimation.

**Figure 6.** Offset in timing between the positive (upper plot) and negative (lower plot) ZDC, were studied with beam splash data from both directions. The timing for the ZDC plus before any corrections show an average time differences that is 16 ns later than ZDC minus.

*3.2. Relative gain verification*

The channel-to-channel response uniformity impacts on the energy resolution. Extensive data taking with splash beam muons resulted in rough verification of the tower's gain equalization. The muons produced from the dumping of the beam passed through the entire detector, depositing large signals in every tower of the ZDC. Typical pulse height distributions of the signal together with a Landau fit are shown in Figures 7 and 8.

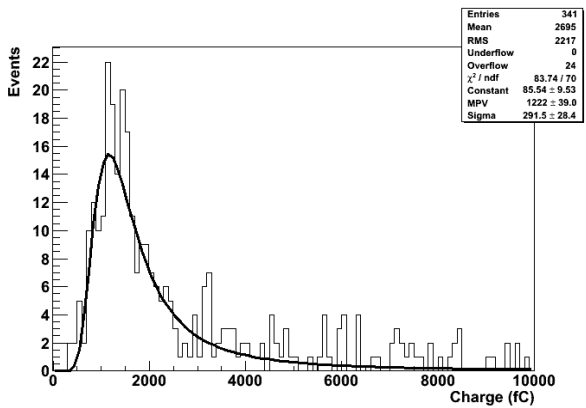
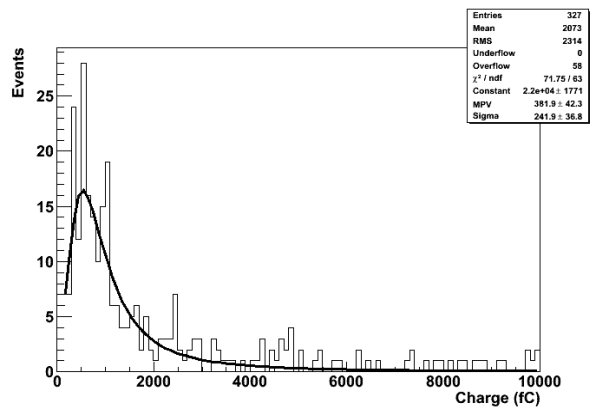

**Figure 7.** Pulse height distribution of energy deposition in one longitudinal readout tower of HAD section.

**Figure 8.** Pulse height distribution of energy deposition in one transverse tower of EM section.

Comparison of the most probably values of Landau fits of tower responses are used to estimate the channel-to-channel gain variation for EM (plus and minus) and HAD (plus and minus) sections separately. After this first set of measurements, we slightly corrected gain (changed of PMT high voltage) of two towers, one in EM and one in HAD. The second set of measurements showed that the relative gain variation (for EM and HAD sections separately) is in range of ~15%. More precise gain measurements and corrections will be done using events from beam-beam collisions.

## 4. Conclusion

The first running experience and commissioning of ZDC with splash events from the first LHC circulating beams were successful. All channels are working as designed. Initially dead - reckoning was used to set up the read-out timing. Beam splash data allowed this timing to be confirmed and optimized. The channel-to-channel gain variation was measured using beam splash events and corrected for uniform response.